\documentclass[superscriptaddress,preprint,aps,prc]{revtex4-1}

\begin{document}

\title{EMMI Rapid Reaction Task Force Meeting on\\ 
``Quark Matter in Compact Star''}

\author{Michael Buballa}
\affiliation{Institut f\"ur Kernphysik (Theoriezentrum), 
Technische Universit\"at Darmstadt, D-64289 Darmstadt, Germany}

\author{Veronica Dexheimer}
\affiliation{Department of Physics, Kent State University, 
Kent, OH 44242, USA}

\author{Alessandro Drago}
\affiliation{Dip.\ di Fisica e Scienze della Terra dell'Universit\`a di Ferrara and
INFN Sez.\ di Ferrara, I-44100 Ferrara, Italy}

\author{Eduardo Fraga}
\affiliation{Institut f\"ur Theoretische Physik, Goethe Universit\"at, 
D-60438 Frankfurt am Main, Germany}
\affiliation{Frankfurt Institute for Advanced Studies, Goethe University, 
D-60438 Frankfurt am Main, Germany}
\affiliation{Instituto de F\`isica, Universidade Federal do Rio de Janeiro, 
Rio de Janeiro, RJ, Brazil} 

\author{Pawel Haensel}
\affiliation{N. Copernicus Astronomical Center, Polish Academy of Sciences, 
Bartycka 18, PL-00-716 Warszawa, Poland}

\author{Igor~Mishustin}
\affiliation{Frankfurt Institute for Advanced Studies, Goethe University, 
D-60438 Frankfurt am Main, Germany}

\author{Giuseppe Pagliara}
\affiliation{Dip.\ di Fisica e Scienze della Terra dell'Universit\`a di Ferrara and
INFN Sez.\ di Ferrara, I-44100 Ferrara, Italy}

\author{J\"urgen Schaffner-Bielich}
\affiliation{Institut f\"ur Theoretische Physik, Goethe Universit\"at, 
D-60438 Frankfurt am Main, Germany}

\author{Stefan Schramm}
\affiliation{Frankfurt Institute for Advanced Studies, Goethe University, 
D-60438 Frankfurt am Main, Germany}

\author{Armen Sedrakian}
\affiliation{Institut f\"ur Theoretische Physik, Goethe Universit\"at, 
D-60438 Frankfurt am Main, Germany}

\author{Fridolin Weber}
\affiliation{Department of Physics, San Diego State University,
San Diego, CA 92182-1233, USA}
\affiliation{Center for Astrophysics and Space Sciences, University of California, 
San Diego, La Jolla, CA 92093, USA}

\begin{abstract}
The recent measurement of two solar mass pulsars has initiated an intense
discussion on its impact on our understanding of the high-density matter in
the cores of neutron stars. A task force meeting was held from October 7-10,
2013 at the Frankfurt Institute for Advanced Studies to address 
the presence of quark matter in these massive stars. During this
meeting, the recent observational astrophysical data and heavy-ion data
was reviewed. The possibility of pure quark stars, hybrid stars and
the nature of the QCD phase transition were discussed and their
observational signals delineated. 
\end{abstract}

\maketitle

\section{Motivation}

The recent observations of the pulsars PSR J1614-2230 and PSR
J0348+0432 proofs beyond doubt the existence of neutron stars with
masses close to two solar masses
\cite{Demorest:2010bx,Antoniadis:2013pzd}. Moreover, there are hints
that there might exist even more massive neutron stars, e.g. from a
recent analysis of the black-widow pulsar PSR J1311-3430
\cite{Romani:2012rh} and pulsar PSR J1816+4510 \cite{Kaplan:2013hii}.

In addition, future astrophysical instruments will set much more
refined constraints on the properties of neutron stars. The future
Five hundred meter Aperture Spherical Telescope FAST and the Square
Kilometre Array SKA will be sensitive enough to find an order of
magnitude more millisecond pulsars and pulsars in binary systems than
those found up to now, increasing the number of pulsar mass
measurements accordingly \cite{Smits:2009vd,Cordes:2005gp}. The
gravitational wave detectors advanced LIGO and advanced VIRGO, both
starting operation in 2014, will have an order of magnitude increase
in their sensitivity to find gravitational waves from compact binary
mergers with a projected rate of measuring about several events per
year \cite{Abadie:2010cf}. Such increased sensitivity will allow
scientists to determine the basic properties of binary neutron star
systems to within a few percent \cite{Aasi:2013jjl}. Finally, future
X-ray satellite missions such as NICER \cite{NICER} and LOFT
\cite{Feroci:2011jc,Bozzo:2011vn,Bozzo:2013txa} plan to measure the radius of
compact stars to unprecedented precision.

In reaction to these astrophysical observations and future prospects,
it seems timely to discuss the role of quark matter in the cores of
neutron stars. We do that by investigating quark matter properties
that are relevant for compact stars and the possible impact of quark
matter on astrophysical observations. A rapid reaction task force
meeting of the ExtreMe Matter Institute (EMMI) on ``Quark Matter in
Compact Stars'' was held from October 7-10, 2013 at the Frankfurt
Institute for Advanced Studies (FIAS) in Frankfurt/Main,
Germany. During the task force meeting, the properties of dense quark
matter relevant for compact star physics were discussed in intense
closed working sessions. A previous meeting of this kind focussing on
the role of hyperons in neutron stars and the problem of low maximum
masses of compact stars in microscopic models with hyperons,
Hyperons2012, was held in Warsaw in November of 2012.

Some of the key questions addressed in the task force meeting on quark matter
in compact stars were
\begin{itemize}

\item what are the properties of dense matter in the core of massive
  pulsars with two solar masses?

\item can one constrain effective models of dense QCD matter?

\item is there still room for quark matter to be present in neutron stars?
 
\item would the observation of a say 2.4 solar mass neutron star
  exclude exotic matter in its core?

\item is there a smoking gun for the presence of quark matter in compact stars?

\item are quark stars still a viable possibility? Can quark stars and
  hadronic stars coexist?

\item what can we learn from radius measurements about the properties of QCD
  matter at extreme densities?

\end{itemize}
The theoretical investigation of these questions is related to two
main research themes: the structure and dynamics of neutron star
matter, and the properties of the quark-gluon plasma and the phase
structure of strongly interacting matter. The properties of dense
matter in the core of a neutron star determine the global properties of
compact stars, i.e.\ the mass-radius relation, which was a key
issue of the task force meeting. This is also relevant for the
investigation of the QCD phase diagram and its possible phase
transitions at high densities and low temperatures. Possible cross
relations of these two research areas were unavoidably part of the
scientific discussion of the task force meeting. The presentations are
available at the task force website at 
https://indico.gsi.de/conferenceTimeTable.py?confId=2420\#all.detailed.

\section{Quark Matter in Compact Stars}

\subsection{Observational Astrophysical Data}

Measurements of pulsar masses in binaries provide most valuable
information on the underlying equation of state (EoS) of star matter
because they, being deduced from binary system parameters, are model
independent within a given theory of gravity. An example of this is
the recent discovery of a compact star with a mass of $1.97 M_\odot$,
which sets an observationally ``clean'' lower bound on the maximum
mass of a compact star via the measured Shapiro delay
\cite{Demorest:2010bx}, spurred an intensive discussion of the phase
structure of dense matter in compact stars consistent with this
limit. The Shapiro delay was detected in the data of
Ref.~\cite{Demorest:2010bx} with extremely high significance. In this
work, robust error estimates were obtained using a Markov chain Monte
Carlo (MCMC) approach to explore effectively all fit dimensions
simultaneously. From the Shapiro delay, the companion white dwarf mass
was deduced to be $0.500\pm 0.006 M_\odot$, which suggests that the
companion is a helium-carbon-oxygen white dwarf.  The fits to the data
also suggest that the binary system is highly edge-on with respect to
the observer with an inclination of $89.17^\circ \pm 0.02^\circ$.  The
amplitude of the Shapiro delay increases strongly with increasing
binary inclination. The fortunate combination of the high orbital
inclination and massive white dwarf companion result in a Shapiro
delay amplitude which is orders of magnitude larger than for most
other millisecond pulsars with smaller mass companions (typically
$0.1-0.2 M_\odot$) and smaller inclinations angles.  The knowledge of
the companion mass combined with the standard Keplerian orbital
parameters and the assumption that the binary is correctly described
by general relativity determine the pulsar’s mass to be $1.97 \pm 0.04
M_\odot$.

Unfortunately, the Shapiro delay mechanism does not provide
information on the radius of the pulsar.  Pulsar radii have been
extracted, e.g., from modeling the X-ray binaries under certain
reasonable model assumptions, but the uncertainties are large. A
recent example is the pulse phase-resolved X-ray spectroscopy of PSR
J0437-4715, the nearest known millisecond pulsar
\cite{Bogdanov:2012md}. The continuum emission has at least three
thermal components; the highest-temperature radiation is naturally
originating from the hot magnetic polar caps of the neutron star,
whereas the low-temperature emission components originates from the
bulk of the surface. The pulse phase-resolved X-ray spectroscopy of
PSR J0437-4715 sets a lower limit on the radius of a $1.76 M_\odot$
solar mass compact star of $R > 11$~km within a $3\sigma$ error. 

A second measurement of a massive compact star was reported in April
2013 by Antoniadis et al.  \cite{Antoniadis:2013pzd}. The measurement
method differs from the Shapiro delay mechanism and thus provides a
model-independent confirmation of the existence of massive compact
stars. To measure the mass of the pulsar, researchers studied in
detail the spectrum of the companion white dwarf using observational
data from the VLT. The measured radial velocity of the companion
combined with the orbital fit, provided the ratio of the masses of the
pulsar and the white dwarf to be $11.70 \pm 0.13$. Further, the mass
of the white dwarf was deduced by modeling the cooling evolution,
which yields for the mass of the white dwarf $0.165-0.185 M_\odot$ at
99.73\% confidence. From the obtained white dwarf mass and the
observed mass ratio $q$, one then finds that the neutron star mass is
in the range of $1.97-2.05 M_\odot$ at 68.27\%, or $1.90-2.18 M_\odot$
at 99.73\% confidence. The authors also measured the orbital decay
$\dot P_b.$ Given this quantity, the value of $q$, and assuming that
General Relativity (GR) is the correct theory of gravity, they
obtained for the pulsar mass $2.07\pm 0.20 M_\odot$ at 68.27\%
confidence. Furthermore, the error in this measurement could be
narrowed down substantially within the next couple of years using
improved measurements of $\dot P_b$.

Larger than two solar mass pulsars were claimed to be observed with
lower confidence in black-widow systems
\cite{vanKerkwijk:2010mt,Romani:2012rh}. Van Kerkwijk et al.\ claim
that, most likely, PSR B1957+20 is a massive neutron star, with a mass
of $2.40 M_\odot$. Their results depend among other things on the
lightcurve modelling of the companion, which is based on a number of
conservative assumptions, as for example, no heat transport over the
surface of the star. From their conservative constraints on both the
inclination of the orbits and the mass ratio, the lower limit on the
pulsar mass is $1.66M_\odot$, well below the massive pulsars mentioned
above. Romani et al.\ \cite{Romani:2012rh} carried out spectroscopic
observations of PSR J1311-3430, which is a gamma-ray black-widow
pulsar. Their modeling of the quiescent light curve provides
constrains on the orbital inclination and mass of the pulsar and its
companion. The most simplistic light-curve fits give $M = 2.7
M_\odot$. Their multicomponent fits imply a large mass range, however
all solutions give pulsar masses in excess of $2.1 M_\odot$.

\subsection{Quark phase transition in compact stars, 
supernovae and gamma ray bursts} 

Several scenarios of the quark phase transition in compact stars have
been discussed in the literature.  They can roughly be grouped into
two categories, depending on when the phase transition takes place
\cite{Drago:2008tb}:
\begin{itemize}
\item quark phase transition before deleptonization of the proto-neutron star
\item quark phase transition during or after the deleptonization.
\end{itemize}
In the first case, the phase transition from hadronic to quark matter
could already occur in the early postbounce phase of a core-collapse
supernova
\cite{Gentile:1993ma,Drago:1997tn,Nakazato:2008su,Sagert:2008ka,Fischer:2010wp,Sagert:2011fq}.
One feature of this mechanism is that it requires a particularly soft
EoS, since the formation of a mixed phase of quarks and hadrons has to
take place at the relatively low densities reached at the moment of
core bounce (or immediately after, during the fallback but anyway
before deleptonization). Since the densities reached at the moment of
the bounce are only moderately dependent on the mass of the
progenitor, this mechanism is rather “universal”, affecting most of
the supernovae, although its effect on the explosion can still depend
on the mass of the progenitor. In this case, a signal in the emitted
neutrinos can be detected \cite{Dasgupta:2009yj}, associated with the
second shock wave which forms when the critical density, separating
the mixed phase and the pure quark matter phase, is reached.  This
neutrino flux would provide a unequivocal signature of a phase
transition associated with the SN explosion. Unfortunately, at the
moment it seems not possible to reconcile the criterion of a phase
transition close to the SN bounce, with a correspondingly soft EoS,
with the lower bound on the neutron star maximum mass, which requires a
hard EoS at high densities \cite{Sagert:2011fq}.

The second possibility is that the quark phase transition takes places
only after, at least partial, deleptonization
\cite{Pons:2001ar,Aguilera:2002dh}. In fact, it is well known, that
when the pressure due to leptons decreases, the central baryonic
density increases and, therefore, hyperon production and the
transition to quark matter proceeds easier. A crucial point in this
analysis is that the quark phase transition can take place only when
strangeness (either in the form of hyperons or kaons) starts being
produced in the center of the compact star: at that point, a
transition to strange quark matter is possible on the time-scale of
the strong interaction \cite{Iida:1998pi}. Clearly, a temporal
separation can exist in this second scenario, between the moment of
core collapse and the moment in which quark matter appears. In this
scenario it is possible to have a huge release of energy associated
with the combustion of hadronic matter into quark matter
\cite{Bombaci:2000cv}. This energy can be the source that powers the
central engine of Gamma-Ray Bursts
\cite{Berezhiani:2002ks,Drago:2004vu}. The time-delay between the
moment of the SN explosion and the moment of the quark phase
transition could explain a few observed features of Gamma-Ray Bursts,
as e.g.\ the existence of very long quiescent times seen in a few
bursts \cite{Drago:2005rc} and the possible existence of Gamma-Ray
Bursts for which no associated SN explosion is observed
\cite{Dermer:2007tj}.

\subsection{Quark Matter Equation of State}

The equation of state for strongly interacting matter at finite
density, i.e.\ the one that should be extracted from QCD in a dense
medium, is a key ingredient for the determination of the structure of
compact stars. Currently, we are still far from a robust calculation on
the lattice in this regime. Nevertheless, one can use extreme limits
in baryonic density, where either QCD can be solved in some
approximation scheme or nuclear theory models become reliable.

At low densities, below and at the nuclear saturation density,
relativistic and non-relativistic nuclear theory descriptions have
achieved a high degree of sophistication and can also be tested by
various means. The problem arises in the adjacent region of high
densities, where quark matter could also be present. So far, most
treatments that include quark matter in compact stars, or are meant to
exclude it, still employ the bag model
\cite{Witten:1984rs,Alcock:1986hz,Haensel:1986qb}.  This corresponds
to the crudest (yet originally efficient, in the vacuum) way of
incorporating confinement, which plainly discards interactions (except
from eventual simplified $O(\alpha_{s})$ correction). However, after
more than a decade of lattice QCD studies at finite temperature and
zero density, it became clear that this description is very
poor. Besides missing the nature of the phase transition, it also
misses the fact that interactions are non-negligible even well above
the critical region of deconfinement.

A widely used alternative to bag-model equations of state is the
Nambu--Jona-Lasinio (NJL) model, which has been applied to compact
stars for about 15
years~\cite{Schertler:1999xn,Baldo:2002ju,Buballa:2003et,Klahn:2006iw}.
An attractive feature of this kind of model is that the ``bag'' pressure
(i.e., the pressure difference between perturbative and non-trivial
vacuum) is not an input parameter but is dynamically generated via
spontaneous chiral-symmetry breaking.  This mechanism also leads to
effective (``constituent'') quark masses, which are (as the bag
pressure) density and temperature dependent.  The same is true for the
pairing gaps in various color-superconducting phases, which can
be incorporated naturally \cite{Buballa:2003qv,Anglani:2013gfu}.

On the other hand, the NJL model is non-renormalizable and therefore
its application is restricted to temperatures and chemical potentials
well below the cutoff scale.  In fact, except for the trivial
Stefan-Boltzmann limit in the case without vector interactions, its
asymptotic behavior does not agree in general with perturbative QCD.
Unfortunately, the standard mean-field approximation does not allow
for a realistic description of the hadronic phase either. Although
interesting attempts have been made to employ the NJL model beyond
mean field in order to construct nuclear matter starting from quark
degrees of freedom~\cite{Lawley:2006ps,Wang:2010iu}, the results are
not yet competitive with state-of-the-art nuclear equations of state.
NJL models are therefore mostly used to model the quark-matter
equation of state above, but not too far above the deconfinement phase
transition, where non-perturbative effects (in particular in the
strange-quark and color superconducting sector) still play a decisive
role.  Of course, it is not clear whether model parameters, which have
been fitted to vacuum quantities, are still valid in this regime.  One
may also question the identification of the NJL vacuum pressure with
the pressure of the hadronic vacuum, so that an additional bag
constant could be introduced~\cite{Pagliara:2007ph}.  Therefore, while
the NJL model can give interesting qualitative insights and uncover
new effects, its quantitative predictions should be taken with care.

In that vein, although perturbation theory at finite density should
only be reliable at densities much higher than those expected to be
found in compact stars, it is a clean framework that contains the
effects from interactions and also provides an estimate of its
uncertainties in a systematic way, via the dependence on the scale
associated with the subtraction point for renormalization. This kind
of control is hardly possible in the case of the bag model
approach. Moreover, perturbation theory for dense QCD corresponds to
the limit any effective treatment should reach at high enough values
of the baryonic chemical potential.

The thermodynamic potential for cold QCD was first computed within
perturbation theory to $\sim\alpha_s^2$ decades ago
\cite{Freedman:1976xs,Freedman:1976dm,Freedman:1976ub,Freedman:1977gz,Baluni:1977ms,Toimela:1984xy}.
More than ten years ago, corrections $\sim\alpha_s^2$ with a modern
definition of the running coupling constant were used to model the
non-ideality in the equation of state for cold, dense QCD with
massless quarks \cite{Fraga:2001id}. This description also predicted
quark stars with masses above 2$M_{\odot}$. (For higher-order
computations, see Refs. \cite{Vuorinen:2003fs,Ipp:2003yz,Ipp:2006ij}).
Remarkably, results from the simple perturbative approach and from
treatments that resort to resummation methods and quasiparticle model
descriptions
\cite{Peshier:1999ww,Peshier:2002ww,Blaizot:2000fc,Andersen:2002jz,Rebhan:2003wn}
seem to agree well even for quark chemical potentials $\mu_q \sim
1~$GeV and smaller.

More recently, it was shown that the inclusion of the running strange quark
mass has a non-trivial influence on the equation of state \cite{Fraga:2004gz}
(see also Refs. \cite{Alford:2002rj,Alford:2004pf}), where it was argued that
effects from both the color superconductivity gap and the strange quark mass
should matter in the lower-density sector of the equation of state. Moreover,
${\cal O}(\alpha_s^2)$ calculations including one massive quark flavor
allowing for further non-perturbative effects found that perturbation theory
converges reasonably well for quark chemical potentials above $1$ GeV
\cite{Kurkela:2009gj}.  The latter also suggest that quark matter in compact
star cores becomes confined to hadrons only slightly above the density of
atomic nuclei, and obtain hybrid star masses of up to $M\sim 2M_{\odot}$, and
even larger in the case of strange stars \cite{Kurkela:2009gj,Kurkela:2010yk}.

\subsection{Quark--Hybrid Stars}

Strange quark matter has been hypothesized to exist in self-bound
(strange) quark stars, but may also be found in the cores or ordinary
``neutron'' stars due to the enormously high pressure values at the
centers of such objects. The latter are referred to as quark-hybrid
stars (see \cite{Weber:2004kj} and references therein). These types of
stars are comprised of a quark core surrounded by a hadronic shell,
which can also contain strangeness via hyperons or a kaon
condensate. Usually, the modeling for these stars consists of
employing two equations of state, a hadronic and a quark one,
switching from one set of degrees of freedom to the other when the
equations of state cross. Common choices for the quark EoS are based
on the bag model, linear $\sigma$ model, non-linear realization of the
$\sigma$ model, NJL model, or Schwinger-Dyson approaches.

The Maxwell and the Gibbs constructions of the phase transition are standard
choices. In the second case, a mixed phase of quarks and hadrons
appears \cite{Glendenning:1992vb,Na:2012td}. Which of the scenarios
applies depends strongly on the QCD surface tension, whose value is
still highly uncertain. Calculations that include macroscopic
structures known as pasta phases show that small surface tensions
yield a behavior similar to the Gibbs construction, whereas larger
surface tensions reproduce the results of the Maxwell construction
\cite{Yasutake:2012dw}, reducing the softening of the EoS due to the
phase mixing.

A crucial ingredient for obtaining heavy hybrid stars with masses of
two $M_\odot$ or larger is a stiff equation of state for the
quarks. While the simple bag model tends to be too soft to sustain a
large core of quark or mixed matter, additional repulsive forces
between the quarks prevent a large jump in energy density (latent
heat) at the transition \cite{Coelho:2010fv,Lenzi:2012xz} and lead to
the stabilization of heavy hybrid stars. In this case, complex stars
containing nucleons, hyperons and various color superconducting phases
are possible \cite{Bonanno:2011ch,Weissenborn:2011qu}.
Frequently, reconfinement can occur at high chemical potentials,
reverting the system to a hadronic one. This can be avoided by taking
into account the finite size of the hadron and including a moderate
excluded volume \cite{Schramm:2013pma}. We note that an open question
remains, however, regarding the discrepancy with lattice results (for
the pressure at low baryonic chemical potential) when quark-vector
interactions are included \cite{Steinheimer:2010sp}.

A signature of the existence of hybrid stars could come from a twin
star solution, in which two stars (a hadronic and a hybrid one) can
exist as stable solutions with similar mass but distinctly different
radii (see
\cite{Gerlach:1968zz,Kampfer:1981yr,Glendenning:1998ag,Schertler:2000xq,Alvarez-Castillo:2013cxa,Negreiros:2010tf,Schramm:2013pma}). 
Such scenarios could, in principle, be validated as soon as
simultaneous radius and mass measurements become available and they
would also point towards a first-order phase transition
at small temperatures. A further possible signature of hybrid stars
would be a distinct frequency range of gravitational wave emission in
the g-mode \cite{Flores:2013yqa}, pulsar backbending
\cite{Glendenning:1997fy} and anomalies in the spin-up of neutron
stars in low-mass X-ray binaries \cite{Glendenning:2000zz}.

In the case that the interiors of high mass compact stars contain
quark matter in a crystalline color superconducting state
\cite{Lin:2007rz,Knippel:2009st,JohnsonMcDaniel:2012wg} or solid quark
matter \cite{Xu:2003xe} a detectable gravitational wave emission is
expected. The characteristic strain of gravitational wave emission
predicted by these models could be observed with advanced LIGO. Such a
detection can potentially put constraints on the breaking strain and
pairing gap of the crystalline color superconducting matter.

Whereas the aforementioned studies assume a first-order transition
from hadrons to quarks, also a continuous cross-over is possible
\cite{Bratovic:2012qs}, similar to the transition at zero chemical
potential. This can be done in a unified description of quarks and
hadrons and leads to a hybrid star with a continuous increase of the
quark fraction without Gibbs construction, generating heavy hybrid
stars \cite{Schramm:2013pma}.

Quark matter in high-mass neutron stars has very recently been
studied using extensions of the local and non-local Nambu-Jona Lasinio
model supplemented with repulsive vector interactions among the quarks
(see \cite{Orsaria:2012je,Orsaria:2013hna} and references therein). The phase
transition from hadronic matter to quark matter
has been constructed via the Gibbs condition, which imposes global
rather than local electric charge neutrality and baryon number
conservation. Depending on the strength of the quark vector repulsion,
it was found that an extended mixed phase of hadronic matter and
quarks can exist in neutron stars as massive as $2.1$ to
$2.4\, M_\odot$. A phase of pure quark matter inside such high-mass
neutron stars, while not excluded, is only obtained for certain
parametrizations of the underlying Lagrangians.  The radii of all these
stars are between 12 and 13~km, as expected for neutron stars of that
mass.

Finally, at extremely large chemical potentials, not necessarily
present inside compact stars, the hadronic or hybrid (containing a
mixed phase) EoS must connect with the perturbative QCD EoS. As shown
in \cite{Kurkela:2009gj,Kurkela:2010yk}, this is not a simple task,
since the perturbative calculation presents error bands and it stops
being reliable for baryon chemical potentials of about 1 GeV. In
addition, so far, we cannot determine in a model-independent fashion
if the hadronic or hybrid EoS will connect to the pure quark one at
low or high densities.  A further difficulty arises when chiral
symmetry is considered. The restoration of chiral symmetry is an
important characteristic of matter at high densities, but it does not
necessarily coincide with the critical density for the deconfinement
\cite{Borsanyi:2010bp,Bazavov:2011nk}. Such an assumption results, for
example, in the quarkyonic matter picture
\cite{McLerran:2007qj,Andronic:2009gj}.

Recently, an unprecedented fast cooling of the compact star in
Cassiopeia A (Cas A) -- the youngest known supernova remnant in Milky
Way -- was inferred from the Chandra observations over a period of ten
years \cite{Elshamouty:2013nfa}. The age of this object is 330 yr. The theoretical
fits to the cooling behaviour of Cas A can provide important
information on the internal composition of compact stars.  Hadronic
models of compact stars have been invoked to fit the data by assuming
a canonical 1.4 $M_{\odot}$ mass neutron star.  The rapid cooling in
these models are mainly due to the pair-breaking process in the
hadronic core of the neutron star
\cite{Page:2010aw,Blaschke:2011gc,Negreiros:2011ak}. An alternative
model of the cooling of the compact star in Cassiopea A is based
on the modeling of the cooling of hybrid stars and accounts for its
unusually fast cooling behavior by invoking a phase transition
among two different color superconducting phases of
quark matter \cite{Sedrakian:2013pva}.  Specifically, the rapid cooling may be
interpreted as an enhancement in the neutrino emission triggered by a
transition from a fully gapped, two-flavor, color-superconducting
phase to a crystalline phase or an alternative gapless,
color-superconducting phase. By fine-tuning a single parameter –- the
temperature of this transition -– a specific cooling scenario can be
selected that fits the Cas A data. Such a scenario requires a massive
$M \sim 2M_{\odot}$ star and, if correct, should be one of the signals
of quark matter in compact stars.

\subsection{Dynamics of Transitions from Nucleonic to Hybrid Stars}

Within the hypothesis of absolutely stable strange quark matter (the
Bodmer-Witten hypothesis) there is the possibility that neutron stars
convert into quark stars \cite{Itoh:1970uw}. Once a seed of strange
quark matter is formed within hadronic matter, the conversion proceeds
as a combustion process where hadronic matter represents the fuel and
quark matter the ash. There have been many works on this subject
indicating that strong deflagration is the relevant combustion mode
for neutron star's burning and that hydrodynamical instabilities play
an important role (see for instance
\cite{Horvath:1988nb,Lugones:1994xg,Drago:2005yj} and references
therein). Recent 3+1D hydrodynamic simulations
\cite{Herzog:2011sn,Pagliara:2013tza} have confirmed that
Rayleigh-Taylor instabilities take place, which significantly increase
the velocity of conversion: almost the whole star is converted in a
time scale of the order of a few milliseconds.  The newly born strange
star is quite hot, with temperatures of a few tens of MeV in the
center and cools down via neutrino diffusion. For the first time the
neutrino signal associated with the conversion has been estimated by
use of a diffusion code \cite{Pagliara:2013tza}. Similarly to the
signal of proto-neutron stars, a proto-quark star would have an
initial luminosity of about $10^{52}$ erg/s, representing thus a
detectable neutrino source if located close enough to our Galaxy.

There are many interesting points that have been raised in the
meeting, the most important concerning the full conversion of the
star: the burning indeed stops before the whole star is converted,
leaving an amount of unburned hadronic material of the order of $0.5
M_{\odot}$.  How the conversion proceeds and how long it takes for a
full conversion to occur, represent important questions for future
studies. It has been also proposed that this result would imply that,
even if the Bodmer-Witten hypothesis is correct, only hybrid stars can
form. This possibility would rule out the existence of pure quark
stars and, possibly, the existence of strangelets in cosmic rays
\cite{Madsen:2004vw}.  On the other hand, the existence of pure quark
stars would imply that, at least in some cases and for some
parametrizations of the equation of state the merger of two strange
stars would eject strangelets \cite{Bauswein:2008gx}, which in turn
would pollute the whole Galaxy potentially triggering the conversion
of all neutron stars \cite{MedinaTanco:1996je} to strange stars. A
better theoretical understanding of the formation of (pure) quark
stars is therefore mandatory if the scenario of the existence of two
families of compact stars \cite{Drago:2013fsa}, hadronic and quark
stars, ought to be viable.

Another interesting question for future studies concerns the possible
transition from deflagration to detonation in the burning of a neutron
star as proposed for instance in Ref.~\cite{Niebergal:2010ds}.  At the moment,
there are no calculations or numerical simulations which confirm this
hypothesis, but this possibility can not be excluded.  A detonation
would be responsible for a significant amount of matter ejected after
the conversion and possible consequences for supernovae and
nucleosynthesis have been analysed (see Ref.~\cite{Ouyed:2013sra}).

Another scenario leading to the transformation of a neutron star into
a hybrid star was also discussed during the meeting. It could be
realized if the phase transition from the hadronic to the quark phase
is strong enough, so that the jump in relative densities exceeds a
critical value of about 1.5
\cite{Ramsey:1950,Lighthill:1950,Kaempfer:1981a}. If in the course of
the evolution (by cooling, slowing down, increasing the mass due to
accretion) the central density reaches this critical value, the star
``rolls down'' into a new equilibrium state, where the quark core
occupies a significant fraction of the star. Up to now only very
schematic calculations of this process have been made, see
\cite{Migdal:1979je,Dimmelmeier:2009vw}. They show that the core
radius first grows exponentially with a characteristic time of a few
milliseconds and then experiences damped oscillations around a new
equilibrium state. As estimated in
Refs.~\cite{Haensel:1982zz,Mishustin:2002xe}, the energy released in
this transition could be as large as $10^{52}$ erg. It should reheat
the star and induce an additional neutrino burst. Certainly, more
realistic simulations of such a catastrophic process could be done
using modern computational resources.

\subsection{Signals for Quark Matter in Compact Stars}

In addition to the signals mentioned above, strange quark stars or the
presence of a new form of matter in the core of compact stars can lead
to the following astrophysical signatures.

\subsubsection{Dressed strange quark stars and Eddington Limit}

A bare quark star differs qualitatively from a neutron star which has
a density at the surface of about 0.1 to 1 g/cm$^3$. The thickness of
the quark surface is just about 1 fm, the length scale of the strong
interaction. The electrons at the surface of a quark star are held to
quark matter electrostatically, and the thickness of the electron
surface is several hundred fermis.  Since neither component, electrons
and quark matter, are held in place gravitationally, the Eddington
limit to the luminosity that a static surface may emit does not apply,
so that bare quark stars may have photon luminosities much greater
than $10^{38}$~erg/s. It has been shown in \cite{Usov:1997ff} that this
value may be exceeded by many orders of magnitude by the luminosity of
$e^+ e^-$ pairs produced by the Coulomb barrier at the surface of a
hot strange star. For a surface temperature of $\sim 10^{11}$~K, the
luminosity in the outflowing pair plasma was calculated to be as high
as $\sim 3 \times 10^{51}$~erg/s.  Such an effect may be a good
observational signature of bare strange stars
\cite{Usov:1997ff,Usov:2001gj,Usov:2001sw,Cheng:2003hv}. If the strange star is
dressed, that is, enveloped in a nuclear crust, however, the surface
made of ordinary atomic matter would be subject to the Eddington
limit. Hence the photon emissivity of a dressed quark star would be
the same as for an ordinary neutron star. But note that if quark matter at the
stellar surface is in the CFL (Color-Flavor Locked) phase, the process
of $e^+ e^-$ pair creation at the stellar quark matter surface may be
turned off.  This may be different for the early stages of a very hot
CFL quark star \cite{Vogt:2003ph}.

\subsubsection{Mass-Radius relationship and rapid rotation of quark stars}

In contrast to neutron stars, the radii of self-bound quark stars
decrease the lighter the stars, according to $M \propto R^3$. The
existence of nuclear crusts on quark stars changes the situation
drastically \cite{Glendenning:1994zb,Weber:2004kj}.  Since the crust
is bound gravitationally, the mass-radius relationship of quark stars
with crusts is qualitatively similar to neutron stars.  In general,
quark stars with or without nuclear crusts possess smaller radii than
neutron stars. This implies that quark stars have smaller mass
shedding (break-up) periods than neutron stars. Moreover, due to the
smaller radii of quarks stars, the complete sequence of quark
stars -- and not just those close to the mass peak, as it is the case
for neutron stars -- can sustain extremely rapid rotation
\cite{Glendenning:1994zb,Weber:2004kj}.  In particular, a strange star
with a typical pulsar mass of around 1.45~$M_\odot$ has a Kepler
period in the approximate range of $0.55\lesssim{P_{\rm K}}/{\rm
  msec}\lesssim 0.8$ \cite{Glendenning:1994zb,Glendenning:1992}. This
is to be compared with ${P_{\rm K}} \sim 1~{\rm msec}$ for neutron
stars of the same mass.

Another novelty of the strange quark matter hypothesis concerns the
existence of a new class of white-dwarf-like objects, referred to as
strange (quark matter) dwarfs \cite{Glendenning:1994zb}. The mass-radius
relationship of the latter differs somewhat from the mass-radius
relationship of ordinary white dwarfs, which may be testable
observationally in the future. Until recently, only rather vague tests
of the theoretical mass-radius relationship of white dwarfs were
possible. This has changed dramatically because of the availability of
new data emerging from the Hipparcos project
\cite{Provencal:1998zz}. These data allow the first accurate measurements
of white dwarf distances and, as a result, establishing the
mass-radius relation of such objects empirically.

\subsubsection{Features of electrically charged  strange quark stars}

One of the most amazing features of strange quark stars concerns the
existence of ultra-high electric fields on their surfaces, which, for
ordinary (i.e., non-superconducting) quark matter, is around
$10^{18}$~V/cm.  If strange matter forms a color superconductor, as
expected for such matter, the strength of the electric field may
increase to values that exceed $10^{19}$~V/cm. The energy density
associated with such huge electric fields is on the same order of
magnitude as the energy density of strange matter itself, which alters
the masses and radii of strange quark stars at the 15\% and 5\% level,
respectively \cite{Negreiros:2009fd}. Similar effects are predicted
also for hybrid stars with a small net electric charge
\cite{Brillante:2014lwa}.  Such mass increases facilitate the
interpretation of massive compact stars, with masses of around
2~$M_\odot$, as possible strange quark stars.

The electrons at the surface of a quark star are not necessarily in a
fixed position but may rotate with respect to the quark matter star
\cite{PicancoNegreiros:2010uc}. In this event magnetic fields can be generated
which, for moderate effective rotational frequencies between the
electron layer and the stellar body, agree with the magnetic fields
inferred for several Central Compact Objects (CCOs). These objects
could thus be interpreted as quark stars whose electron atmospheres
rotate at frequencies that are moderately different ($\sim 10$~Hz)
from the rotational frequency of the quark star itself.

Last but not least, we mention that the electron surface layer may be
strongly affected by the magnetic field of a quark star in such a way
that the electron layer performs vortex hydrodynamical oscillations
\cite{Xu:2011ss}. The frequency spectrum of these oscillations has been
derived in analytic form in \cite{Xu:2011ss}. If the thermal X-ray
spectra of quark stars are modulated by vortex hydrodynamical
oscillations, the thermal spectra of compact stars, foremost central
compact objects (CCOs) and X-ray dim isolated neutron stars (XDINSs),
could be used to verify the existence of these vibrational modes
observationally. The central compact object 1E 1207.4-5209 appears
particularly interesting in this context, since its absorption
features at 0.7 keV and 1.4 keV can be comfortably explained in the
framework of the hydro-cyclotron oscillation model \cite{Xu:2011ss}.

\subsubsection{Possible connection of quark stars to SGRs, AXPs, and XDINs}

Rotating superconducting quark stars ought to be threaded with
rotational vortex lines, within which the star's interior magnetic
field is at least partially confined.  The vortices (and thus magnetic
flux) would be expelled from the star during stellar spin-down,
leading to magnetic re-connection at the surface of the star and the
prolific production of thermal energy. In has been shown in
\cite{Niebergal:2009yb} that this energy release can re-heat quark
stars to exceptionally high temperatures, such as observed for Soft
Gamma Repeaters (SGRs), Anomalous X-Ray pulsars (AXPs), and X-ray dim
isolated neutron stars (XDINs), and that SGRs, AXPs, and XDINs may be
linked ancestrally \cite{Niebergal:2009yb}.

\subsubsection{Rotation-driven compositional changes}

The change in central density of a neutron star whose frequency varies
from zero to the mass shedding (Kepler) frequency can be as large as
50 to 60\% \cite{Weber:2004kj}. This suggests that changes in the
rotation rate of a neutron star may drive phase transitions and/or
lead to significant compositional changes in the star's core
\cite{Weber:2004kj,Weber:2006ep,Negreiros:2011ak}. As a case in point,
for some rotating neutron stars the mass and initial rotational
frequency may be just such that the central density rises from below
to above the critical density for dissolution of baryons into their
quark constituents. This may be accompanied by a sudden shrinkage of
the neutron star, effecting the star's moment of inertia and, thus,
its spin-down behavior.  As shown in \cite{Glendenning:1997fy}, the
spin-down of such a neutron star may be stopped or even reversed for
tens of thousands to hundreds of thousands of years
\cite{Weber:2004kj,Glendenning:1997fy}.  The observation of an
isolated neutron star which is spinning-up, rather than down, could
thus hint at the existence of quark matter in
its core.

\subsubsection{Quark-hadron Coulomb lattice  (electron Bremsstrahlung)}

Because of the competition between the Coulomb and the surface
energies associated with the positively charged regions of nuclear
matter and negatively charged regions of quark matter, the mixed phase
may develop geometrical structures (e.g., blobs, rods, slabs),
similarly to what is expected of the sub-nuclear liquid-gas phase
transition \cite{Glendenning:1992vb}.  The consequences of such a Coulomb
lattice for the thermal and transport properties of neutron stars have
been studied in \cite{Na:2012td}.  It was found that at low temperatures
of $T\lesssim 10^8$~K the neutrino emissivity from electron-blob
Bremsstrahlung scattering is at least as important as the total
contribution from all other Bremsstrahlung processes (such as
nucleon-nucleon and quark-quark Bremsstrahlung) and modified nucleon
and quark Urca processes.  It is also worth noting that the scattering
of degenerate electrons off rare phase blobs in the mixed phase region
lowers the thermal conductivity by several orders of magnitude
compared to a quark-hadron phase without geometric patterns.  This may
lead to significant changes in the thermal evolution of the neutron
stars containing solid quark-hadron cores, which has not yet been
studied.

\subsubsection{Pycnonuclear reaction rates}

The presence of strange quark nuggets in the crustal matter of neutron
stars could be a consequence of Witten's strange quark matter
hypothesis. The impact of such nuggets on the pycnonuclear reaction
rates among heavy atomic nuclei has been studied in \cite{Golf:2009jb}.
Particular emphasis was put on the consequences of color
superconductivity on the reaction rates.  Depending on whether or not
quark nuggets are in a color superconducting state, their electric
charge distributions differ drastically, which was found to have
dramatic consequences for the pycnonuclear reaction rates in the
crusts of neutron stars. Future nuclear fusion network calculations
may thus have the potential to shed light on the existence of strange
quark matter nuggets and on whether they are in a color
superconducting state, as suggested by QCD.

\subsubsection{R-mode instability}

The r-mode instability of a rotating neutron star dissipates the
star's rotational energy by coupling the angular momentum of the star
to gravitational waves
\cite{Andersson:1997xt,Lindblom:1998wf,Friedman:1997uh}. This instability can
be active in a newly formed isolated neutron star as well as in old
neutron stars being spun up by accretion of matter from binary
stars. If the interior contains quark matter, the r-mode instability
and the gravitational wave signal may carry information about quark
matter \cite{Jaikumar:2008kh,Rupak:2010qg,Rupak:2012wk,Sad:2008gf,Mannarelli:2008je}. 

\subsubsection{Quark Novae}

The conversion of a neutron star to a hypothetical quark star could
lead to quark novae. Such events could explain gamma ray bursts
\cite{Staff:2007fr}, the production of heavy elements such as platinum
through r-process nucleosynthesis \cite{Jaikumar:2006qx}, and
double-humped super-luminous supernovae \cite{Ouyed:2012mg}.

\section{Relation to other Fields}

\subsection{Hyperons in Compact Stars}

A previous task force meeting called {\bf HyperoNS2012} was held
November 21--24, 2012 at the Nicolaus Copernicus Astronomical Center
(CAMK), in Warsaw, Poland. The scientific program of the meeting was
prepared by Pawel Haensel (CAMK) and J\"urgen Schaffner-Bielich (then
Heidelberg University). The motivation for this task force meeting was
clear: the measurement of a $2M_\odot$ neutron star
\cite{Demorest:2010bx} forced us, dense matter theorists, to
reconsider the problem of the structure of massive neutron star
cores. In particular, the problem of too strong a softening of the
equation of state of dense baryonic matter due to the presence of
hyperons at high density, inconsistent with such a high neutron star
mass, reappeared in a very dramatic way.  Two basic ``solutions'' for
this puzzle were proposed: (a) additional repulsion between hyperons
due to a vector meson coupled to hyperons only; (b) a transition to a
very stiff quark plasma before the hyperon threshold. Note, that both
``solutions'' have their own problems and difficulties.

It seemed timely and useful to organize, two years after the
2$M_\odot$ neutron star announcement, a three-day task force meeting
devoted to the present status of possible existence of hyperon/quark
cores in neutron star. The knowledge of the hyperon interactions was
reviewed. Proponents of different solutions to the ``hyperon puzzle''
had the opportunity to present their arguments, which were then
critically discussed by the participants of the meeting.

Marcello Baldo (Catania) presented a review of the modern many-body
theory of nuclear matter, and its relation to the``hyperon puzzle''. A
non-linear relativistic mean-field (RMF) model yielding neutron star
with hyperon cores and $M_{\rm max}>2M_\odot$ was presented by Ilona
Bednarek (Katowice). David Blaschke (Wroclaw) gave an overview of
quark-matter theories in the context of hybrid stars. Interesting new
results on the appearance of strangeness in neutron star cores were
presented by Micaela Oertel (Meudon). The effect of the symmetry
energy of nuclear matter on the strangeness content of neutron star
cores was discussed by Constanca Providencia (Coimbra). J\"urgen
Schaffner-Bielich (Heidelberg) gave an overview of the antikaon
properties in dense matter. Various approaches to the properties of
hybrid stars, and perspectives for a unified formulation of the
description of hadron and quark phases were described by Stefan
Schramm (Frankfurt). Nonrelativistic Brueckner-Hartree-Fock
calculations of hypernuclear matter with Nijmegen NY potentials were
discussed and their implications for hypernuclei and neutron star were
critically examined by Hans-Josef Schulze (Catania). Armen Sedrakian
(Frankfurt) discussed the structure of color superconducting cores of
hybrid stars, and the cooling of such stars, and confronted
theoretical models with recent observations of cooling of the pulsar
in Cas A. A hybrid model that combines an extended RMF description of
hyperonic matter and a modified Nambu-Jona-Lasinio model for
quark matter applied to hybrid stars was presented by
Stefan Typel (Darmstadt). Isaac Vidana (Coimbra) presented recent
results on the stiffening effect of three-body forces on the EoS of
hypernuclear matter. He showed that the three-body forces are unable
to make the EoS of hyperonic matter sufficiently stiff to yield
$M_{\rm max}>2M_\odot$. Constraints imposed on the phase transition to
quark matter in neutron star cores, implied by the $M_{\rm
  max}>2M_\odot$ condition, were presented by Pawel Haensel (Warsaw).

The talks were followed by lively and inspiring discussion. The two
solar mass limit is found to be a challenge for hyperonic cores in
many-body theories, and a strong tuning following a selection of a RMF
model, is required to find ``a way out''; Brueckner-type theories are
unable to yield $M_{\rm max}>2M_\odot$.  Quark cores replacing a
hyperon core have to be very stiff and the hadron-quark matter phase
transition should be associated with a small density jump very
different from the twin star condition. Several projects were inspired
in these discussions and reported a year later at the EMMI Rapid
Reaction Task Meeting in Frankfurt in October of 2013. In this way the
Warsaw task force initiated a series of meetings on current problems
in the description of compact stars and dense matter physics.

There were 22 participants in the Warsaw meeting. All living and
travel expenses of non-Polish participants (15) were covered by the
CAMK from the COMPSTAR-POL Polish grant.  More details, including
talks from the ``HyperoNS2012'' meeting can be found at
http://www.camk.edu.pl/conf/Hyperons2012.

The hyperon puzzle was also a topic at the task force meeting reported
here during the discussions as it is important to fix the low to
medium density equation of state for hybrid stars. More information on
the hyperon-hyperon interactions is needed to address the issue in
more detail but might be available in the near future. Two particle
correlations of hyperons emitted in relativistic heavy-ion collisions
might come from experiments at CERN's LHC. Recent lattice simulations
of the binding energy of dibaryons with hyperons have already
demonstrated that bound dibaryon states with hyperons might
exist. These lattice results are encouraging but are only calculated
for unphysical (heavy) pion masses and need to be extrapolated to the
physical limit.

\subsection{Messages from Heavy-Ion Physics}

Heavy-ion collisions at intermediate energies bring important information
about properties of compressed baryonic matter. Using ions with different Z/N
ratios opens the possibility to study isospin asymmetric matter. As
calculations show, see e.g.\ Ref.~\cite{Li:2002yda}, baryon densities up to
$(2-4)\rho_0$ can be reached at intermediate stages of heavy-ion collisions at
beam energies of 0.2--2.0~AGeV. This density interval spans a significant
fraction of the interior of compact stars.  Such experiments allow to obtain
valuable information about the density-dependence of the symmetry energy and
the composition of isospin-asymmetric matter. Recently, the old FOPI/LAND data
on elliptic flow of neutrons and light charged particle production in Au+Au
collisions at 0.4 AGeV \cite{Leifels:1993ir} were reanalyzed to extract the
exponent $\gamma$ in the density dependence of the symmetry energy
$(\rho/\rho_0)^\gamma$: comparison of these data with predictions of two
Quantum Molecular Dynamics models \cite{Russotto:2011hq,Cozma:2011nr} has
shown that $\gamma=0.9\pm 0.4$. The analysis of various sets of nuclear data,
sensitive to the density dependence of symmetry energy, has allowed to extract
the value of its first derivative at $\rho\approx \rho_0$, expressed as
$L/(3\rho_0)$. The parameter $L$ is found to lie in the range of about
$40-60$~MeV \cite{Lattimer:2012xj}. 

The analysis of the subthreshold production of kaons in heavy-ion
collisions measured by the KaoS collaboration \cite{Sturm:2000dm}
revealed that the bulk nuclear equation of state around $(2-3)\rho_0$
is soft within transport simulations
\cite{Fuchs:2000kp,Hartnack:2005tr}. Implications for the maximum mass
of neutron stars have been addressed in \cite{Sagert:2011kf} showing
that the equation of state extracted from these analysis is compatible
with the new pulsar mass constraint if the equation of state stiffens
considerably at higher density. Neutron star masses of more than $2.7
M_\odot$ could not be reached in \cite{Sagert:2011kf} when enforcing
causality by using the stiffest possible equation of state and the
constraint from the KaoS data analysis.

Presently, it is well established experimentally that strange hadrons are
abundantly produced in heavy-ion collisions at intermediate energies. Their
observed abundances closely follow the predictions of simple thermal models,
see e.g.~\cite{BraunMunzinger:2001mh,Cleymans:1999st}. In particular, the
ratio of Lambda and proton yields, $\Lambda/p$, at freeze-out is about 1 at
low SPS energies. Transport calculations show that strangeness (and
anti-strangeness) is mostly produced at the initial stage of the collision and
then decreases slowly towards freeze-out \cite{Buss:2011mx}. Thus, one should
conclude that the abundance of strange baryons in the dense baryonic matter
produced in such collisions is comparable to the abundance of nucleons. No
signs of hyperon suppression have been found so far. Therefore, there are no
reasons to ignore the presence of hyperons in dense baryonic matter, as done
in some works. The participants of the task force meeting have agreed that the
models using entirely nucleonic degrees of freedom up to $(4-5)\rho_0$ and
ignoring hyperons should be considered as ``exotic scenarios'', even if they
satisfy the mass-radius constraints on compact stars.

\section{Outlook}

The need for a simultaneous measurement of the mass and the radius to
say a few percent precision seems to be crucial to learn more about
the properties of high-density matter in the cores of compact
stars. Arguments are put forward during the discussions that exotic
matter inside neutron stars implies radii of more than 11 km for a
$1.4M_\odot$ neutron star assuming a speed of sound similar to the one
present in the MIT bag model and a maximum mass of $2M_\odot$ (see
figure 5 of ref.~\cite{Lattimer:2012nd}). As it became clear during
the meeting, the simple MIT bag model is ruled out from lattice data,
so that the arguments have to be taken with great caution. The limit
on the radius was shown to be beaten by more advanced models as
presented during the task force meeting, see e.g.\
\cite{Blaschke:2014via}, so that radii of less than 11 km are possible
with strange matter cores. Still, the issue is not fully addressed and
needs to be reconsidered with improved models for the high-density
equation of state.

It is promising that a new technique has been developed in order to extract
masses and radii of compact stars: the study of the X-ray flux modulations due
to rotating hot spots on the surface of neutron stars. Taking into account
general relativity corrections, the far-side spot becomes more visible for
smaller stars through gravitational light-bending, which depends on the
compactness $M/R$ of the compact star. This technique can be applied to study
thermal pulsation of nearby millisecond pulsars and burst oscillations from
low-mass X-ray binaries.  Both the approved NICER mission of NASA \cite{NICER}
and the proposed LOFT satellite \cite{Feroci:2011jc} are adopting this
technique. Error bars smaller than $\sim$10$\%$ are expected for the radius
measurement, what would finally put rather tight constraints on the dense
matter EoS.

Furthermore, the study of the high-density equation of state for
compact stars seems to be at a stage where the old models of QCD need
to be abandoned. It emerged from the task force that the simple MIT
bag model is excluded from observations and from lattice data. Vector
interactions stiffen the quark matter equation of state but are in
variance with asymptotic freedom and lattice data.  Improved models of
QCD are available but need to be extended for astrophysical
applications to achieve progress in further studies.  Indeed, the task
force meeting succeeded in achieving this task by initiating joint
work where perturbative QCD calculations at high baryon densities were
used to describe the equation of state suitable for compact stars
\cite{Fraga:2013qra,Kurkela:2014vha}. Judging from the lively atmosphere created at
the task force meeting, more joint work in this direction will surely
follow.

\acknowledgments

This work was supported by the Alliance Program of the Helmholtz Association
(HA216/EMMI). We thank all the participants for the lively discussions
and the constructive atmosphere generated at this meeting. 

\bibliography{all_new,myliterature}
\bibliographystyle{utphys}

\end{document}